\documentclass[aps,pra,twocolumn,showpacs]{revtex4-1}

\usepackage{graphicx}

\begin{document}

\title{What the complex joint probabilities observed in weak measurements\\
can tell us about quantum physics}

\author{Holger F. Hofmann}
\email{hofmann@hiroshima-u.ac.jp}
\affiliation{
Graduate School of Advanced Sciences of Matter, Hiroshima University,
Kagamiyama 1-3-1, Higashi Hiroshima 739-8530, Japan}
\affiliation{JST, CREST, Sanbancho 5, Chiyoda-ku, Tokyo 102-0075, Japan
}

\begin{abstract}
Quantum mechanics does not permit joint measurements of non-commuting observables. However, it is possible to measure the weak value of a projection operator, followed by the precise measurement of a different property. The results can be interpreted as complex joint probabilities of the two non-commuting measurement outcomes. Significantly, it is possible to predict the outcome of completely different measurements by combining the joint probabilities of the initial state with complex conditional probabilities relating the new measurement to the possible combinations of measurement outcomes used in the characterization of the quantum state. We can therefore conclude that the complex conditional probabilities observed in weak measurements describe fundamental state-independent relations between non-commuting properties that represent the most fundamental form of universal laws in quantum physics.
\end{abstract}


\maketitle

In physics, measurements should tell us all we need to know about the reality of physical objects. It is therefore extremely perplexing that quantum measurement fails to do so. At the very heart of this failure lies the uncertainty principle: although the individual outcomes of precise measurements appear real enough, such measurements cannot provide us with a satisfactory characterization of the relation between non-commuting properties. 
A possible solution to the problem of measurement uncertainty may be the application of weak measurements, where the measurement interaction is reduced to negligible levels to avoid any disturbance of the final result by the measurement back-action \cite{Aha88}. The weak values obtained in such measurements can describe the correlations between the non-commuting properties of a quantum system and address questions that seemed to be fundamentally inaccessible in conventional quantum measurements.

Recently, a number of experiments have shown that quantum paradoxes can be explained in terms of negative conditional probabilities observed in the weak measurement of the projection operators that are used to represent probabilities in the Hilbert space formalism \cite{Res04,Wil08,Lun09,Yok09,Gog11}. These measurements suggest that a microscopic explanation of quantum paradoxes is possible if the complex conditional probabilities observed in weak measurements are recognized as a fundamental feature of quantum physics. In this short paper, I will give an outline of the fundamental physics described by the generally complex conditional probabilities observed in weak measurement by explaining how the outcomes of non-commuting quantum measurements are related to each other.

The starting point of this explanation is provided by quantum tomography. In fact, the complex probabilities observed in weak measurements were first discovered in the early days of quantum mechanics by Kirkwood \cite{Kir33} and by Dirac \cite{Dir45}, in attempts to formulate a quantum mechanical equivalent to classical phase space. From these early formulations, it is clear that the complex joint probability $\rho(a,b)$ that can be reconstructed from a weak measurement of $\mid a \rangle\langle a \mid$ followed by a precise measurement of $\mid b \rangle$ provides a complete description of the quantum state \cite{Joh07,Hof12a,Lun12}. Specifically, the joint probability is determined from the weak values by multiplication with the post-selection probability and reads
\begin{equation}
\rho(a,b) = \langle b \mid a \rangle \langle a \mid \hat{\rho} \mid b \rangle. 
\end{equation}
The reconstructed density operator is given by 
\begin{equation}
\hat{\rho} = \sum_{a,b} \rho(a,b) \frac{\mid a \rangle \langle b \mid}{\langle b \mid a \rangle}. 
\end{equation}
Interestingly, weak measurements are not the only method of observing these joint probabilities in a fairly direct and unambiguous experiment - they also appear in the correlations between optimally clones systems \cite{Hof12b} and in variable strength measurements with sufficiently controlled back-action \cite{Suz12,Hof12c}.

Once the complex joint probability is determined, it can be used to predict the outcomes of other measurements. The probability of a measurement outcome $m$ represented by a pure state $\mid m \rangle$ is given by
\begin{equation}
\langle m \mid \hat{\rho} \mid m \rangle = \sum_{a,b} \rho(a,b) \frac{\langle b \mid m \rangle \langle m \mid a \rangle}{\langle b \mid a \rangle}. 
\end{equation}
This sum corresponds to Bayes rule for conventional probabilities, where the conditional probability $p(m|a,b)$ is given by the weak value 
\begin{equation}
p(m|a,b) = \frac{\langle b \mid m \rangle \langle m \mid a \rangle}{\langle b \mid a \rangle}. 
\end{equation}
This weak value can be obtained in a completely separate measurement on the initial state $\mid a \rangle$, followed by a post-selection of $\mid b \rangle$. Since it is completely independent of the state $\rho$ and applies to any joint probability $\rho(a,b)$, the complex conditional probability $p(m|a,b)$ is a fundamental law of physics expressing the universal relation between $m$ and $(a,b)$.  

As pointed out in \cite{Hof12a}, the conditional probability also describes transformations between the $(a,b)$ representation and the $(a,m)$ representation of complex joint probabilities. Since these transformations are reversible, the complex conditional probabilities $p(m|a,b)$ do not describe any statistical noise, but should be understood as deterministic descriptions of the relation between $m$, $a$, and $b$. In classical phase space, these three properties would be described by contours that either intersect at one point, so that $m=f_m(a,b)$ is a well-defined function of $(a,b)$, or else enclose a triangular area defined by an action $S(a,b,m)$. As shown in \cite{Hof11}, this action defines the complex phase of the weak conditional probabilities,
\begin{equation}
S(a,b,m) = \hbar \; \mbox{Arg}\left(p(m|a,b)\right).
\end{equation}
Specifically, this phase defines a unitary transformation in $m$ that transforms $a$ into a state that maximally overlaps with $b$. In this sense, the complex phase of $p(m|a,b)$ describes the action-distance of the transformation from $a$ to $b$ along $m$. Since this phase is also the origin of quantum paradoxes, it may be appropriate to think of it as the quantum measure of logical tension between $a$, $b$, and $m$. 

In conclusion, the complex probabilities observed in weak measurement provide a complete description of quantum statistics, where complex conditional probabilities define the fundamental state-independent relations between different physical properties. The phases of complex probabilities describe transformational distances between the physical properties in terms of the action, where a phase of $2 \pi$ corresponds to an action of $h=2 \pi \hbar$. It may therefore be possible to re-formulate quantum physics in terms fundamental laws of physics that describe the relations between different physical properties in terms of their complex conditional probabilities.

This work was supported by JSPS KAKENHI Grant Number 24540427.

\end{document}